\documentclass[12pt]{article}
\usepackage{amssymb}

\def\hybrid{\topmargin 0pt      \oddsidemargin 0pt
        \headheight 0pt \headsep 0pt
        \voffset=-0.5cm
        \hoffset=-0.25in
        \textwidth 6.75in
        \textheight 9.5in       
        \marginparwidth 0.0in
        \parskip 5pt plus 1pt   \jot = 1.5ex}
\catcode`\@=11
\def\marginnote#1{}

\newcount\hour
\newcount\minute
\newtoks\amorpm
\hour=\time\divide\hour by60 \minute=\time{\multiply\hour by60
\global\advance\minute by-\hour}
\edef\standardtime{{\ifnum\hour<12 \global\amorpm={am}%
        \else\global\amorpm={pm}\advance\hour by-12 \fi
        \ifnum\hour=0 \hour=12 \fi
        \number\hour:\ifnum\minute<10 0\fi\number\minute\the\amorpm}}
\edef\militarytime{\number\hour:\ifnum\minute<10 0\fi\number\minute}

\def\draftlabel#1{{\@bsphack\if@filesw {\let\thepage\relax
   \xdef\@gtempa{\write\@auxout{\string
      \newlabel{#1}{{\@currentlabel}{\thepage}}}}}\@gtempa
   \if@nobreak \ifvmode\nobreak\fi\fi\fi\@esphack}
        \gdef\@eqnlabel{#1}}
\def\@eqnlabel{}
\def\@vacuum{}
\def\draftmarginnote#1{\marginpar{\raggedright\scriptsize\tt#1}}
\def\draftlabel#1{{\@bsphack\if@filesw {\let\thepage\relax
   \xdef\@gtempa{\write\@auxout{\string
      \newlabel{#1}{{\@currentlabel}{\thepage}}}}}\@gtempa
   \if@nobreak \ifvmode\nobreak\fi\fi\fi\@esphack}
        \gdef\@eqnlabel{#1}}
\def\@eqnlabel{}
\def\@vacuum{}
\def\draftmarginnote#1{\marginpar{\raggedright\scriptsize\tt#1}}

\def\draft{\oddsidemargin -.5truein
        \def\@oddfoot{\sl preliminary draft \hfil
        \rm\thepage\hfil\sl\today\quad\militarytime}
        \let\@evenfoot\@oddfoot \overfullrule 3pt
        \let\label=\draftlabel
        \let\marginnote=\draftmarginnote
   \def\@eqnnum{(\theequation)\rlap{\kern\marginparsep\tt\@eqnlabel}%
\global\let\@eqnlabel\@vacuum}  }

\def\numberbysection{\@addtoreset{equation}{section}
        \def\theequation{\thesection.\arabic{equation}}}

\def\underline#1{\relax\ifmmode\@@underline#1\else
        $\@@underline{\hbox{#1}}$\relax\fi}

\def\titlepage{\@restonecolfalse\if@twocolumn\@restonecoltrue\onecolumn
     \else \newpage \fi \thispagestyle{empty}\c@page\z@
        \def\thefootnote{\fnsymbol{footnote}} }

\def\endtitlepage{\if@restonecol\twocolumn \else  \fi
        \def\thefootnote{\arabic{footnote}}
        \setcounter{footnote}{0}}  

\numberbysection \hybrid

\newcounter{mo}

\newcommand{\tr}{{\rm tr}}
\newcommand{\ti}[1]{\tilde{#1}}

\newcommand{\de}{\delta}

\newcommand{\Mat}{ {\rm Mat}(N,\mathbb C) }

\newcommand{\mC}{\mathbb C}
\newcommand{\mZ}{\mathbb Z}

\newcommand{\sgn}{\hbox{sign}}

\newcommand{\e}{\varepsilon}
\newcommand{\h}{\hbar}

\newcommand{\n}{\eta}
\newcommand{\tL}{{\tilde L}}

\newtheorem{predl}{Proposition}[section]

\def\beq{\begin{equation}}
\def\eq{\end{equation}}
\def\p{\partial}

\def\res{\mathop{\hbox{Res}}\limits}

\begin{document}

\setcounter{page}{1}

\date{}
\date{}
\vspace{50mm}

\begin{flushright}
 ITEP-TH-35/18\\
\end{flushright}
\vspace{0mm}

\begin{center}
\vspace{0mm}
{\LARGE{Trigonometric integrable tops from solutions of}}
 \\ \vspace{4mm}
 {\LARGE{associative Yang-Baxter equation}}
\\
\vspace{12mm} {\large \ \ {T. Krasnov}\,$^{\natural\, \sharp\, \S}$\
\ \ \ \ \ \ \ \ \ \ {A. Zotov}\,$^{\diamondsuit\, \sharp\, \S\,
\natural}$ }
 \vspace{8mm}

\vspace{1mm} $^\diamondsuit$ -- {\small{\rm
 Steklov Mathematical Institute of Russian Academy of Sciences,\\ Gubkina str. 8, Moscow,
119991,  Russia}}\\
 \vspace{1mm} $^\sharp$ -- {\small{\rm 
 ITEP, B. Cheremushkinskaya str. 25,  Moscow, 117218, Russia}}\\
 \vspace{1mm}$^\S$ - {\small{\rm National Research University Higher School of Economics, \\
 Usacheva str. 6,  Moscow, 119048, Russia}}
 \\
  \vspace{1mm} $^\natural$ -- {\small{\rm Moscow Institute of Physics and Technology,\\ Inststitutskii per.  9, Dolgoprudny,
 Moscow region, 141700, Russia}}

\end{center}

\begin{center}\footnotesize{{\rm E-mails:}{\rm
 \ timofei.krasnov@phystech.edu,\
zotov@mi-ras.ru}}\end{center}
%
%

 \begin{abstract}
We consider a special class of quantum non-dynamical $R$-matrices in
the fundamental representation of ${\rm GL}_N$ with spectral
parameter given by trigonometric solutions of the associative
Yang-Baxter equation. In the simplest case $N=2$ these are the
well-known 6-vertex $R$-matrix and its 7-vertex deformation. The
$R$-matrices are used for construction of the classical relativistic
integrable tops of the Euler-Arnold type. Namely, we describe the
Lax pairs with spectral parameter, the inertia tensors and the
Poisson structures. The latter are given by the linear Poisson-Lie
brackets for the non-relativistic models, and by the classical
Sklyanin type algebras in the relativistic cases. In some particular
cases the tops are gauge equivalent to the Calogero-Moser-Sutherland
or trigonometric Ruijsenaars-Schneider models.
 \end{abstract}

\newpage

 \small{
\tableofcontents  
 }


\section{Introduction}
\setcounter{equation}{0}
In this paper we discuss ${\rm GL}_N$ integrable Euler-Arnold type
tops \cite{Arnold} defined by the equations of motion
  \beq\label{q001}
  \dot{S}=[S,J(S)]\,,\qquad S=\sum\limits_{i,j=1}^N E_{ij}S_{ij}\in\Mat\,,
  \eq
where $\{S_{ij}\,,\ i,j=1,...,N\}$ is the set of dynamical
variables, $\{E_{ij}\}$ -- is the standard basis in $\Mat$, and
$J(S)$ is a linear map\footnote{Equations (\ref{q001}) describe
rotation of a rigid body in $N$-dimensional (complex) space. In this
respect $J(S)$ is the inverse inertia tensor. } on $S$
  \beq\label{q002}
  J(S)=\sum\limits_{i,j,k,l=1}^N J_{ijkl}E_{ij}S_{lk}\in\Mat
  \eq
with components $J_{ijkl}$ independent of dynamical variables. The
model is not integrable in the general case but for special choices
of $J(S)$ only.
The construction of integrable tops under consideration goes back to
E. Sklyanin's paper \cite{Skl1} (see also \cite{FT}). The idea was
to formulate the classical analogue of the models described by the
inverse scattering method. In this way the classical spin chains
were described and the quadratic Poisson structures were obtained
via the classical limit of the exchange (RLL) relations.

The ${\rm GL}_N$ top can be viewed as the model obtained through the
classical limit from the 1-site spin chain. The rational models of
this type were described in \cite{AASZ,LOZ8}.
Here we use a specification of the above mentioned results based on
trigonometric $R$-matrices satisfying the associative Yang-Baxter
equation (AYBE) \cite{FK,Pol}:
  \beq\label{q100}
  R^\hbar_{12}(z_{12})
 R^{\eta}_{23}(z_{23})=R^{\eta}_{13}(z_{13})R_{12}^{\hbar-\eta}(z_{12})+
 R^{\eta-\hbar}_{23}(z_{23})R^\hbar_{13}(z_{13})\,,\quad
 z_{ab}=z_a-z_b\,.
  \eq
 It was shown in \cite{LOZ16} that solution of (\ref{q100})
 satisfying also additional properties of skew-symmetry\footnote{$P_{12}$
 in (\ref{q110}) and below is the permutation operator. In particular, for any pair of matrices $A,B\in\Mat$
 with $\mC$-valued matrix elements: $(A\otimes B) P_{12}=P_{12}(B\otimes A)$.}
  \beq\label{q110}
  \begin{array}{c}
  \displaystyle{
 R^\hbar_{12}(z)=-R_{21}^{-\hbar}(-z)=-P_{12}R_{12}^{-\hbar}(-z)P_{12}\,,
 \qquad\qquad
 P_{12}=\sum\limits_{i,j=1}^N E_{ij}\otimes E_{ji}\,,
 }
 \end{array}
 \eq
unitarity
 \beq\label{q120}
   \begin{array}{c}
 \displaystyle{
R^\hbar_{12}(z) R^\hbar_{21}(-z) = f^\hbar(z)\,\,1_N\otimes 1_N
 }
  \end{array}
  \eq
and the local expansions\footnote{Here we imply that $R$-matrices
have simple poles at $z=0$ and $\hbar=0$ only, and no higher order
poles. The classical limit is given by the expansion near $\hbar=0$.
It is the first one condition in (\ref{q140}). See
(\ref{q400})-(\ref{q402}).}
  \beq\label{q140}
  \begin{array}{c}
  \displaystyle{
 \res\limits_{\hbar=0}R^\hbar_{12}(z)=1_N\otimes
1_N=1_{N^2}\,,\quad\quad \res\limits_{z=0}R^\hbar_{12}(z)=P_{12}
 }
 \end{array}
 \eq
($1_N$ -- is $N\times N$ identity matrix) leads to explicit
constructions of the Lax pair $L(z),M(z)\in\Mat$. That is the Lax
equations
  \beq\label{q003}
  {\dot L}(z)=[L(z),M(z)]
  \eq
are equivalent to the equations of motion (\ref{q001}) identically
in spectral parameter $z$. All the data of the models including
their Hamiltonians, the Lax pairs, the Poisson structures and the
inertia tensors (i.e. $J(S)$) are given in terms of coefficients of
expansion of the $R$-matrices near $\hbar=0$ and $z=0$. For example,
in the relativistic case the Lax pair is as follows:
  \beq\label{q004}
  \begin{array}{c}
  \displaystyle{
 L^\eta(z)=\tr_2(R^\eta_{12}(z)S_2)\,,
 \qquad\qquad
 M^\eta(z)=-\tr_2(r_{12}(z)S_2)\,,
 }
 \end{array}
 \eq
 where $S_2=1_N\otimes S$, and $r_{12}(z)$ -- is the classical
 $r$-matrix. See Section \ref{sect3} for details. The Planck constant plays the role of the relativistic
 deformation parameter $\eta$. In some special case it is identified
 with the corresponding parameter in the Ruijsenaars-Schneider
 model.

Notice that together with the properties (\ref{q110}) and
(\ref{q120}) a solution of (\ref{q100}) satisfies also the custom
 Yang-Baxter equation
  \beq\label{q130}
R_{12}^\hbar(z_1-z_2)R_{13}^\hbar(z_1-z_3)R_{23}^\hbar(z_2-z_3)=
R_{23}^\hbar(z_2-z_3)R_{13}^\hbar(z_1-z_3)R_{12}^\hbar(z_1-z_2)\,,
  \eq
so that such solution of (\ref{q100}) is then a true quantum
$R$-matrix by convention.
Sometimes the following property holds true as well\footnote{The
condition (\ref{q150}) is related to the finite Fourier
transformations. See \cite{Z2} for details.}:
  \beq\label{q150}
  \begin{array}{c}
  \displaystyle{
 R^\hbar_{12}(z)P_{12}=R^{z}_{12}(\hbar)\,.
 }
 \end{array}
 \eq
 This allows to relate the coefficients of expansion (of $R$-matrices) near $\hbar=0$ and
$z=0$ with each other.

The paper is organized as follows. In Section \ref{sect2} we
describe the set of well-known trigonometric $R$-matrices satisfying
conditions (\ref{q100})-(\ref{q140}), and briefly describe the
general classification of such solutions of (\ref{q100}) suggested
by T. Schedler and A. Polishchuk \cite{T,Pol2}.
We will show that a representative example of the classification is
given by the so-called non-standard trigonometric $R$-matrix
\cite{AHZ}, which generalizes the ${\rm GL}_2$ 7-vertex $R$-matrix
\cite{Chered} for $N>2$. In Section \ref{sect3} we review the
construction of integrable tops and evaluate all the data for the
general case and the non-standard $R$-matrix. Using (\ref{q100}) we
also prove that the classical quadratic $r$-matrix structure
provides the classical Sklyanin type Poisson structure. This results
in getting the classification of the trigonometric Sklyanin Poisson
structures, and it is parallel to the classification of solutions of
the associative Yang-Baxter equation.
In Section \ref{sect4} we consider a special top corresponding to
rank one matrix $S$ and related to the non-standard $R$-matrix.  It
turns out that this model is gauge equivalent to the
Ruijsenaars-Schneider \cite{Ruijs1} or the Calogero-Moser-Sutherland
\cite{Ca} models. Explicit changes of variables are described




\paragraph{Acknowledgments.} The work
was supported in part by RFBR grant 18-01-00926 and the
 Russian Academy of Sciences program ''Nonlinear dynamics: fundamental
problems and applications''. The research of A. Zotov was supported
in part by the HSE University Basic Research Program, Russian
 Academic Excellence Project '5-100' and by the Young Russian Mathematics award.





\section{Trigonometric $R$-matrices and AYBE}\label{sect2}
\setcounter{equation}{0}

 We begin with the properties of well-known $R$-matrices and then proceed to the
 general case.

\subsection{Standard and non-standard $R$-matrices}



Consider the following examples of $R$-matrices
  \beq\label{q2}
  \begin{array}{c}
  \displaystyle{
 R^{\eta}_{12}(z)=\sum\limits_{i,j,k,l=1}^N R^\eta_{ijkl}(z)E_{ij}\otimes
 E_{kl}
 }
 \end{array}
 \eq

\noindent $\bullet$ \underline{ The $\mZ_N$-invariant $A_{N-1}$
trigonometric $R$-matrix \cite{Chered,Perk,Kulish}}:
  \beq\label{q200}
   \begin{array}{c}
   \displaystyle{
  (R_1)^{\n}_{ij,kl}(z)= \de_{ij}\de_{kl}\de_{ik}\,\frac{N}{2}\,\Big(\coth(Nz/2)+\coth(N\n/2)\Big)+
  }
  \\ \ \\
   \displaystyle{
  +\, \de_{ij}\de_{kl}\e(i\ne k)\frac{N e^{(i-k)\n-\sgn(i-k)N\n/2}}{2\sinh(N\n/2)}+\de_{il}\de_{kj}
  \e(i\ne k)\frac{N e^{(i-k)z-\sgn(i-k)Nz/2}}{2\sinh(Nz/2)}\,,
  }
  \end{array}
  \eq
where hereinafter we use
  \beq\label{q201}
  \begin{array}{c}
  \displaystyle{
 \e(\hbox{A})=\left\{\begin{array}{l} 1\,,\hbox{if A is true}\,,\\ 0\,,\hbox{if A is false}\,.\end{array}\right.
 }
 \end{array}
 \eq

 \vskip2mm
\noindent  $\bullet$   \underline{Baxterization of the
(trigonometric) Cremmer-Gervais $R$-matrix \cite{Babelon,AHZ}}:
  \beq\label{q205}
   \begin{array}{c}
   \displaystyle{
  (R_2)^{\n}_{ij,kl}(z)= \de_{ij}\de_{kl}\de_{ik}\,\frac{N}{2}\,\Big(\coth(Nz/2)+\coth(N\n/2)\Big)+
  }
  \\ \ \\
   \displaystyle{
  +\, \de_{ij}\de_{kl}\e(i\ne k)\frac{N e^{(i-k)\n-\sgn(i-k)N\n/2}}{2\sinh(N\n/2)}+\de_{il}\de_{kj}
  \e(i\ne k)\frac{N e^{(i-k)z-\sgn(i-k)Nz/2}}{2\sinh(Nz/2)}\, +
  }
  \\ \ \\
     \displaystyle{
  + N\de_{i+k,j+l}\Big(\e(i{<}j{<}k) e^{(i-j)z+(j-k)\n}-\e(k{<}j{<}i)
  e^{(i-j)z+(j-k)\n}\Big)\,.
 }
  \end{array}
  \eq
  It differs from the previous one (\ref{q200}) by the last line.
  Let us comment on how it is related to the Cremmer-Gervais $R$-matrix. First, one should
  perform the gauge transformation
  \beq\label{q206}
   \begin{array}{c}
   \displaystyle{
 R_{12}^\eta(z-w)\rightarrow {\ti R}_{12}^\eta(z,w)=D_1(z) D_2(w) R^{\n}_{12}(z)
 D_1^{-1}(z)D_2^{-1}(w)
 }
  \end{array}
  \eq
  with the diagonal matrix $D_{ij}(z)=\delta_{ij}e^{-jz}$. For
  (\ref{q205}) ${\ti R}_{12}^\eta(z,w)={\ti R}_{12}^\eta(z-w)$. The
  result is
  \beq\label{q207}
   \begin{array}{c}
   \displaystyle{
  (\tilde{R}_2)^{\n}_{ij,kl}(z)= \de_{ij}\de_{kl}\de_{ik}\,\frac{N}{2}\,\Big(\coth(Nz/2)+\coth(N\n/2)\Big)+
  }
  \\ \ \\
   \displaystyle{
  +\, \de_{ij}\de_{kl}\e(i\ne k)\frac{N e^{(i-k)\n-\sgn(i-k)N\n/2}}{2\sinh(N\n/2)}+\de_{il}\de_{kj}
  \e(i\ne k)\frac{N e^{\sgn(i-k)Nz/2}}{2\sinh(Nz/2)}\, +
  }
  \\ \ \\
     \displaystyle{
  + N\de_{i+k,j+l}\Big(\e(i{<}j{<}k) e^{(j-k)\n}-\e(k{<}j{<}i)
  e^{(j-k)\n}\Big)\,.
 }
  \end{array}
  \eq
Consider the Cremmer-Gervais $R$-matrix \cite{CG}. It is free of
spectral parameter:
  \beq\label{q208}
   \begin{array}{c}
   \displaystyle{
  R^{\rm CG,q}_{12}=q^{-1/N}\Big(
  q\sum\limits_{i=1}^N E_{ii}\otimes E_{ii}
  +q\sum\limits_{i>j}^N q^{-2(i-j)/N} E_{ii}\otimes E_{jj}
  +q^{-1}\sum\limits_{i<j}^N q^{-2(i-j)/N} E_{ii}\otimes E_{jj}-
  }
  \\ \ \\
   \displaystyle{
 -(q-q^{-1})\sum\limits_{i<j}^N\sum\limits_{k=1}^{j-i-1}q^{2k/N}
 E_{j-k,i}\otimes E_{i+k,j}+
 (q-q^{-1})\sum\limits_{i>j}^N\sum\limits_{k=0}^{i-j-1}q^{-2k/N}
 E_{j+k,i}\otimes E_{i-k,j}\Big)\,.
  }
  \end{array}
  \eq
Next, introduce
  \beq\label{q209}
   \begin{array}{c}
   \displaystyle{
R^{\rm CG,q}_{12}(x)=xR_{12}^{\rm CG,q}-x^{-1}\left(R_{21}^{\rm
CG,q}\right)^{-1}\,.
 }
  \end{array}
  \eq
Finally,
  \beq\label{q2091}
   \begin{array}{c}
   \displaystyle{
(\tilde{R}_2)^{\eta}_{12}(z)=-\frac{N}{4\sinh(Nz/2)\sinh(N\eta/2)}\,R^{\rm
CG,q}_{12}(x)^T\,,
 }
  \end{array}
  \eq
where ''T'' means the transpose of matrix
($R_{ij,kl}\stackrel{T}{\rightarrow}R_{ji,lk}$) and
$x=e^{-\eta/2-Nz/2}$, $q=e^{-N\eta/2}$.

\vskip2mm
 \noindent  $\bullet$  \underline{Non-standard
trigonometric $R$-matrix \cite{AHZ}}:
  \beq\label{q210}
   \begin{array}{c}
   \displaystyle{
  R^{\n}_{ij,kl}(z)= \de_{ij}\de_{kl}\de_{ik}\,\frac{N}{2}\,\Big(\coth(Nz/2)+\coth(N\n/2)\Big)+
  }
  \\ \ \\
   \displaystyle{
  +\, \de_{ij}\de_{kl}\e(i\ne k)\frac{N e^{(i-k)\n-\sgn(i-k)N\n/2}}{2\sinh(N\n/2)}+\de_{il}\de_{kj}
  \e(i\ne k)\frac{N e^{(i-k)z-\sgn(i-k)Nz/2}}{2\sinh(Nz/2)}\, +
  }
  \\ \ \\
     \displaystyle{
  + N\de_{i+k,j+l}\Big(\e(i{<}j{<}k) e^{(i-j)z+(j-k)\n}-\e(k{<}j{<}i)
  e^{(i-j)z+(j-k)\n}\Big)+
  }
  \\ \ \\
     \displaystyle{
  + N
  \de_{i+k,j+l+N}\Big(\de_{iN}e^{-jz-l\n}-\de_{kN}e^{lz+j\n}\Big)\,.
 }
  \end{array}
  \eq
It differs from the previous one (\ref{q205}) by the last line,
which provides in $N=2$ case the 7-vertex deformation \cite{Chered}
of the 6-vertex $R$-matrix.

\vskip3mm

\noindent{\bf Properties of $R$-matrices.}

 Briefly, all the $R$-matrices (\ref{q200}), (\ref{q205}) and (\ref{q210})
 satisfy the associative Yang-Baxter equation (\ref{q100}), the
 skew-symmetry property (\ref{q110}), the unitarity property
 (\ref{q120}), and therefore, the Yang-Baxter equation (\ref{q130}). Moreover, all of them satisfy
  the Fourier symmetry (\ref{q150}). The gauge transformed
  $R$-matrix (\ref{q207}) does not satisfy (\ref{q150}) while the
  rest of the properties hold true.

In order to summarize the properties of the above $R$-matrices
introduce notations for the last lines of (\ref{q205}) and
(\ref{q210}): $\Delta_1R^{\n}(z)=(R_2)^{\n}(z)-(R_1)^{\n}(z)$ and
$\Delta_2R^{\n}(z)=(R)^{\n}(z)-(R_2)^{\n}(z)$, i.e.
  \beq\label{q212}
   \begin{array}{c}
   \displaystyle{
  \Delta_1R^{\n}_{ij,kl}(z)=
  N\de_{i+k,j+l}\Big(\e(i{<}j{<}k) e^{(i-j)z+(j-k)\n}-\e(k{<}j{<}i)
  e^{(i-j)z+(j-k)\n}\Big)\,,
 }
  \end{array}
  \eq
  \beq\label{q213}
   \begin{array}{c}
   \displaystyle{
  \Delta_2R^{\n}_{ij,kl}(z)=
  N
  \de_{i+k,j+l+N}\Big(\de_{iN}e^{-jz-l\n}-\de_{kN}e^{lz+j\n}\Big)
 }
  \end{array}
  \eq
and consider the following linear combination:
  \beq\label{q214}
   \begin{array}{c}
   \displaystyle{
  {\bf R}^\eta(z)=A_0(R_1)^\eta(z)+A_1\Delta_1R^{\n}(z)+A_2\Delta_2R^{\n}(z)\,,
 }
  \end{array}
  \eq
where $A_0$, $A_1$ and $A_2$ are some constants. For example, for
$A_0=A_1=A_2=1$ (\ref{q214}) yields (\ref{q210}). To summarize:

\begin{predl}
For any $A_0$, $A_1$ and $A_2$ (\ref{q214}) satisfies the properties
(\ref{q110}), (\ref{q150}) and (\ref{q120}) with
  \beq\label{q215}
   \begin{array}{c}
   \displaystyle{
  f^\eta(z)=A_0^2\frac{N^2}{4}\left(\frac{1}{\sinh^2(N\eta/2)}-\frac{1}{\sinh^2(Nz/2)}\right)\,,
 }
  \end{array}
  \eq
  that is (\ref{q214}) is non-degenerated iff $A_0\neq 0$.

The associative Yang-Baxter equation (\ref{q100}) holds true for all
$R$-matrices (\ref{q200}), (\ref{q205}) and (\ref{q210}). The linear
combination
 (\ref{q214}) satisfies (\ref{q100}) in the following cases:

1. $A_0=A_1\neq 0$, $A_2$ -- any,

2. $A_0\neq 0$, $A_1=A_2=0$

3. $A_0=A_1=0$, $A_2$ -- any.

The latter means that the $R$-matrix (\ref{q213}) satisfies
(\ref{q100}).
 \end{predl}

Let us also mention two special cases:

a.) In the case\footnote{In fact, for $N=2$ case $A_1$ is not
necessary since $\Delta_1R^{\n}(z)=0$ in this case.} $N=2,3$ the
combination  (\ref{q214}) satisfies (\ref{q100}) for $A_0$, $A_1$ --
any, and $A_2=0$.

b.) for $N=4$ and $A_0=A_2=0$ (\ref{q214}) does not satisfy
 (\ref{q100}) while the Yang-Baxter equation (\ref{q130}) holds
 true.

 The case 2 from the Proposition can be verified directly.
 Instead of a direct proof of cases 1. and 3. we will show (in the next subsection) that the
 non-standard $R$-matrix (\ref{q210}) is contained in the general
 classification. Next, we can apply the gauge transformation (\ref{q206})
 with
  \beq\label{q218}
   \begin{array}{c}
   \displaystyle{
 D_{ij}=\delta_{ij} e^{-j\Lambda}
 }
  \end{array}
  \eq
to (\ref{q210}). In terms of components it leads to
$R^{\n}_{ij,kl}(z) \to e^{(j+l-i-k)\Lambda}R^{\n}_{ij,kl}(z)$.
Therefore, the last line of (\ref{q210}) is multiplied by
$e^{-N\Lambda}$:
  \beq\label{q2101}
   \begin{array}{c}
   \displaystyle{
  R^{\n}_{ij,kl}(z)= \de_{ij}\de_{kl}\de_{ik}\,\frac{N}{2}\,\Big(\coth(Nz/2)+\coth(N\n/2)\Big)+
  }
  \\ \ \\
   \displaystyle{
  +\, \de_{ij}\de_{kl}\e(i\ne k)\frac{N e^{(i-k)\n-\sgn(i-k)N\n/2}}{2\sinh(N\n/2)}+\de_{il}\de_{kj}
  \e(i\ne k)\frac{N e^{(i-k)z-\sgn(i-k)Nz/2}}{2\sinh(Nz/2)}\, +
  }
  \\ \ \\
     \displaystyle{
  + N\de_{i+k,j+l}\Big(\e(i{<}j{<}k) e^{(i-j)z+(j-k)\n}-\e(k{<}j{<}i)
  e^{(i-j)z+(j-k)\n}\Big)+
  }
  \\ \ \\
     \displaystyle{
  + N
  e^{-N\Lambda}
  \de_{i+k,j+l+N}\Big(\de_{iN}e^{-jz-l\n}-\de_{kN}e^{lz+j\n}\Big)\,.
 }
  \end{array}
  \eq
By taking the limit $\Lambda\rightarrow\pm\infty$ we come to the
cases 1. with $A_2=0$ or to the case 3.

\vskip4mm At last, consider

\noindent $\bullet$  \underline{$R$-matrix for the affine quantized
algebra ${\hat{\mathcal U}}_q({\rm gl}_N)$
 \cite{Jimbo,RTF}}:
  \beq\label{q270}
   \begin{array}{c}
   \displaystyle{
  R^{{\rm xxz,}\n}_{12}(z)=
   \frac{N}{2}\Big(\coth(Nz/2)+\coth(N\n/2)\Big)\sum\limits_{i=1}^N E_{ii}\otimes E_{ii}+
  }
  \\ \ \\
   \displaystyle{
 +\frac{(N/2)}{\sinh(N\eta/2)}\sum\limits_{i\neq j}^N E_{ii}\otimes E_{jj}+
 \frac{(N/2)}{\sinh(Nz/2)}\sum\limits_{i< j}^N
 \left( E_{ij}\otimes E_{ji}\,e^{Nz/2}+E_{ji}\otimes
 E_{ij}\,e^{-Nz/2}\right)\,.
  }
  \end{array}
  \eq
 It is used for construction of ${\rm GL}_N$ XXZ
spin chains and is usually written in different normalization:
  \beq\label{q271}
   \begin{array}{c}
   \displaystyle{
  {\ti R}^{{\rm xxz,}q}_{12}(x)=\frac4N\,\sinh(Nz/2)\sinh(N\eta/2)\,R^{{\rm xxz,}\n}_{12}(z)=
  }
  \\ \ \\
   \displaystyle{
   =\left(xq-\frac{1}{xq}\right)\sum\limits_{i=1}^N E_{ii}\otimes E_{ii}+
 \left(x-\frac1x\right)\sum\limits_{i\neq j}^N E_{ii}\otimes E_{jj}+
 \left(q-\frac1q\right)\sum\limits_{i\neq j}^N x^{\sgn(j-i)}
 E_{ij}\otimes E_{ji}\,,
  }
  \end{array}
  \eq
  where $x=e^{Nz/2}$, $q=e^{N\eta/2}$.
The XXZ $R$-matrix is the Baxterization of the Drinfeld's one
\cite{Dr}:
  \beq\label{q272}
   \begin{array}{c}
   \displaystyle{
  \left({ R}^{{\rm Dr,}q}_{12}\right)^{\pm 1}
   =q^{\pm 1}\sum\limits_{i=1}^N E_{ii}\otimes E_{ii}+
 \sum\limits_{i\neq j}^N E_{ii}\otimes E_{jj}\pm
 (q-q^{-1})\sum\limits_{i>j}^N
 E_{ij}\otimes E_{ji}\,.
  }
  \end{array}
  \eq
Namely,
  \beq\label{q273}
   \begin{array}{c}
   \displaystyle{
  {\ti R}^{{\rm xxz,}q}_{12}(x)=xR^{{\rm Dr,}q}_{21}-x^{-1}\left( R^{{\rm
  Dr,}q}_{12}\right)^{-1}\,.
  }
  \end{array}
  \eq
The $R$-matrix (\ref{q270}) satisfies Yang-Baxter equation
(\ref{q130}). It is skew-symmetric and unitary (\ref{q120}) with
  \beq\label{q274}
   \begin{array}{c}
   \displaystyle{
 f^\eta(z)=\frac{N^2}{4}\left(\frac{1}{\sinh^2(N\eta/2)}-\frac{1}{\sinh^2(Nz/2)}\right)\,.
  }
  \end{array}
  \eq
The associative Yang-Baxter equation (\ref{q100}) for (\ref{q270})
holds true in the $N=2$ case. For $N>2$ the difference of the l.h.s.
and the r.h.s. from (\ref{q100}) is not zero though it is
independent of spectral parameters:
  \beq\label{q275}
   \begin{array}{c}
   \displaystyle{
  R^\hbar_{12}(z_{12})
 R^{\eta}_{23}(z_{23})-R^{\eta}_{13}(z_{13})R_{12}^{\hbar-\eta}(z_{12})-
 R^{\eta-\hbar}_{23}(z_{23})R^\hbar_{13}(z_{13})=
  }
  \\ \ \\
   \displaystyle{
=-\frac{N^2}{8\cosh(N\hbar/4)\cosh(N\eta/4)\cosh(N(\hbar-\eta)/4)}
 \sum\limits^N_{i\neq j\neq k\neq i} E_{ii}\otimes E_{jj}\otimes
 E_{kk}\,.
  }
  \end{array}
  \eq
The latter statement is verified by direct computation. We do not
consider the XXZ $R$-matrix for construction of integrable tops in
this paper. It is of course possible, but our method requires
(\ref{q100}) to be valid.

%

\subsection{General classification}

Here we briefly describe the classification \cite{T,Pol2} of
trigonometric solutions to associative Yang-Baxter equation
(\ref{q100}) with the properties of skew-symmetry (\ref{q110}) and
unitarity (\ref{q120}). As noted previously, this is sufficient
condition for satisfying the Yang-Baxter equation (\ref{q130}) as
well. So that we deal with the quantum non-dynamical $R$-matrices.
 Another goal of the Section is to show how the non-standard
$R$-matrix (\ref{q210}) arises from the classification.

General solution of (\ref{q100}) is given in terms of combinatorial
construction called the associative Belavin-Drinfeld structure.
Consider $S=\{1,...,N\}$ -- a finite set of $N$ elements. Say, $S$
is the set of $N$ vertices on a circle numerated from $1$ to $N$
(the extended Dynkin diagram of $A_{N-1}$ type). Let $C_0$ be a
transitive cyclic permutation acting on $S$, and $\Gamma_{C_0}$ be
its graph, i.e. the set of ordered pairs
$\Gamma_{C_0}=\{(s,C_0(s))\,,s\in S\}$.

Define another one transitive cyclic permutation $C$ and a pair of
proper subsets $\Gamma_1,\Gamma_2\subset\Gamma_{C_0}$ related by
$C$: $(C\times C)\Gamma_1=\Gamma_2$, where the action means
$(C\times C) (i,j)=(C(i),C(j))$. So that $C\times C$ provides the
induced bijective map $\tau$: $\Gamma_1\stackrel{C\times
C}{\longrightarrow}\Gamma_2$. The set $(\Gamma_1,\Gamma_2,\tau)$ is
an example of the Belavin-Drinfeld triple \cite{BD}.

Here the action of $\tau$ is extended to larger sets.
 Namely,
 it is extended to $\tau:\, P_1\stackrel{C\times
C}{\longrightarrow}P_2$, where $P_{1,2}$ are the following sets:
  \beq\label{q301}
   \begin{array}{c}
   \displaystyle{
  P_i=\{ (s,C^k_0(s)):\, (s,C_0(s))\in\Gamma_i\,,
  ...\,,(C^{k-1}_0(s),C^k_0(s))\in\Gamma_i\,, (C_0^k(s),C^{k+1}_0(s))\notin\Gamma_i
   \}\,.
  }
  \end{array}
  \eq
From the transitivity of $C$ and the choice of $\Gamma_{1,2}$ to be
{\em proper} subsets of $\Gamma_{C_0}$ it follows that there exists
a number $k$ such that $(C\times C)^k\Gamma_1\notin\Gamma_1$.
Similarly, there exist $k_1,k_2$ with the property $(C_0\times
C_0)^{k_{i}+1}\Gamma_{i}\notin\Gamma_{i}$, $i=1,2$. Therefore, $P_i$
are well-defined finite sets, and $\tau$ is the bijective map
between them.

Then the \underline{general answer for trigonometric $R$-matrix
based on $(C_0,C,\Gamma_1,\Gamma_2)$} is as follows:
  \beq\label{q302}
   \begin{array}{c}
      \displaystyle{
  R^{\eta}_{12}(z)=\frac{N}{2}\Big(\coth(Nz/2)+\coth(N\n/2)\Big)\sum_{i} E_{ii}\otimes
  E_{ii}+
  }
   \\ \ \\
   \displaystyle{
  +\frac{N}{e^{N\n}-1}\sum\limits_{0<n<N\,,\, i=C^n(k)} e^{n\n}E_{ii}\otimes E_{kk}+
  \frac{N}{e^{Nz}-1}\sum\limits_{0<m<N\,,\,k=C^m_0(i)}   e^{mz} E_{ik}\otimes
  E_{ki}+
  }
  \\ \ \\
   \displaystyle{
  +\sum\limits_{\hbox{\tiny${\begin{array}{c}0<m<N\,,\,n>0,\\
  i=C^m_0(j)\,,\,\tau^n(j,i)=(k,l)\end{array}}$}}
  N
  \left(e^{-n\n-mz}E_{ij}\otimes E_{kl}- e^{n\n+mz}E_{kl}\otimes
  E_{ij}\right)\,,
   }
  \end{array}
  \eq
where the sums are over all possible values of indices -- elements
of $S$. In particular, the last sum is over all
$i,j,k,l\in\{1,...,N\}$ and positive $m,n$ for which the
$\tau^n(j,i)$ is defined, i.e. $(j,i)\in P_1$ and
$\tau^n(j,i)=(k,l)\in P_2$ with $i=C^m_0(j)$.
The $R$-matrix is skew-symmetric and unitary (\ref{q120})  with
$f^\eta(z)$  (\ref{q274}). The answer (\ref{q302}) is given e up to
some gauge transformations. See the  details in \cite{T,Pol2}.

\noindent {\bf Example.} Consider example with the cyclic
permutations
  \beq\label{q303}
  C_0:\quad
   \begin{array}{|c|c|}
   \hline
   \hline
  \ \ s\ \ & C_0(s)
   \\
   \hline
   1 & N
      \\
   \hline
   2 & 1
      \\
   \hline
   3 & 2
      \\
   \hline
   \vdots & \vdots
         \\
   \hline
   N & N-1
   \\
   \hline
  \hline
  \end{array}
  \qquad\qquad\qquad
  C=C_0^{-1}:\quad
   \begin{array}{|c|c|}
   \hline
   \hline
   s & C(s)
   \\
   \hline
   1 & 2
      \\
   \hline
   2 & 3
      \\
   \hline
   \vdots & \vdots
      \\
   \hline
   N-1 & N
         \\
   \hline
   N & 1
   \\
   \hline
  \hline
  \end{array}
  \eq
and the proper subsets
$\Gamma_{1,2}\subset\Gamma_{C_0}=\{(s,C_0(s))$ given by
  \beq\label{q304}
  \begin{array}{c}
      \displaystyle{
\Gamma_1=\Big\{ (1,N)\,, (2,1)\,, (3,2)\,, ... \,, (N-1,N-2)
\Big\}\,,
     }
  \end{array}
  \eq
  \beq\label{q305}
  \begin{array}{c}
      \displaystyle{
\Gamma_2=(C\times C)\Gamma_1=\Big\{ (2,1)\,, (3,2)\,,  ... \,,
(N-1,N-2)\,, (N,N-1) \Big\}\,.
     }
  \end{array}
  \eq
To construct $P_1$ consider the action of $C_0\times C_0$ on the
elements of $\Gamma_1$ (\ref{q304}):
  \beq\label{q306}
  \begin{array}{l}
      \displaystyle{
 (1,N)\,\stackrel{C_0\times C_0}{\longrightarrow}\,
 (N,N-1)\not\subset\Gamma_1\,,
     }
     \\
     (2,1)\,\stackrel{C_0\times C_0}{\longrightarrow}\, (1,N)\,\stackrel{C_0\times C_0}{\longrightarrow}\,
 (N,N-1)\not\subset\Gamma_1\,,
     \\
   (3,2)\,\stackrel{C_0\times C_0}{\longrightarrow}\,
    (2,1)\,\stackrel{C_0\times C_0}{\longrightarrow}\, (1,N)\,\stackrel{C_0\times C_0}{\longrightarrow}\,
 (N,N-1)\not\subset\Gamma_1\,,
 \\
  \vdots
  \\
  (N-1,N-2)\,\stackrel{C_0\times C_0}{\longrightarrow}\,
  \ldots
  \stackrel{C_0\times C_0}{\longrightarrow}\,(2,1)\,\stackrel{C_0\times C_0}{\longrightarrow}\,
  (1,N)\,\stackrel{C_0\times C_0}{\longrightarrow}\,
 (N,N-1)\not\subset\Gamma_1\,.
  \end{array}
  \eq
 According to the definition (\ref{q303}) we get the following set
 for $P_1$:
  \beq\label{q307}
  P_1=\left\{
  \begin{array}{l}
      \displaystyle{
 (1,N)\,,
     }
     \\
     (2,1)\,,\ \ (2,N)\,,
     \\
   (3,2)\,,\ \
    (3,1)\,,\ \ (3,N)\,,
 \\
  \qquad\qquad\vdots
  \\
  (N\!-\!1,N\!-\!2)\,,(N\!-\!1,N\!-\!3)\,,...\,,(N\!-\!1,1)\,,(N\!-\!1,N)
  \end{array}
  \right\}
  \eq
In a similar way from (\ref{q305}) we obtain the set of $P_2$:
  \beq\label{q308}
  \begin{array}{l}
  P_2=(C\times C)P_1=\left\{
  \begin{array}{l}
      \displaystyle{
 (2,1)\,,
     }
     \\
     (3,2)\,,\ (3,1)\,,
     \\
   (4,3)\,,\
    (4,2)\,,\ (4,1)\,,\
 \\
  \qquad\qquad\vdots
  \\
  (N,N\!-\!1)\,,(N,N\!-\!2)\,,...\,,(N,2)\,,(N,1)
  \end{array}
  \right\}
    \end{array}
  \eq
The bijection between $P_1$ and $P_2$ induced by $C\times C$ is the
map $\tau$.

 \begin{predl}
The $R$-matrix (\ref{q302}) reproduces the non-standard one
(\ref{q210}) for the case of the associative Belavin-Drinfeld
structure (\ref{q303})-(\ref{q305}).
 \end{predl}
\noindent\underline{\em{Proof:}}\quad The first lines of
(\ref{q302}) and (\ref{q210}) coincide. Consider the first term from
the second line of (\ref{q302}):
  \beq\label{q310}
  \begin{array}{c}
      \displaystyle{
\frac{N}{e^{N\n}-1}\sum\limits_{0<n<N\,,\, i=C^n(k)}
e^{n\n}E_{ii}\otimes
E_{kk}=\frac{Ne^{-N\eta/2}}{2\sinh(N\eta/2)}\sum\limits_{0<n<N\,,\,
i=C^n(k)} e^{n\n}E_{ii}\otimes E_{kk}
     }
  \end{array}
  \eq
Due to the definition of $C$ (\ref{q303}) for the summation index
$n$ we have: $n=i-k$ if $i>k$ and $n=N-k+i$ for $i<k$. In this way
we reproduce the first term in the second line of (\ref{q210}).
Similar consideration for the second term in the second line of
(\ref{q302}) yields that the total second line of (\ref{q302})
coincides with the second line of (\ref{q210}).

Next, consider the first sum in the last line of (\ref{q302}) and
subdivide it into two parts:
  \beq\label{q311}
  \begin{array}{c}
      \displaystyle{
\sum\limits_{\hbox{\tiny${\begin{array}{c}0<m<N\,,\,n>0,\\
  i=C^m_0(j)\,,\,\tau^n(j,i)=(k,l)\end{array}}$}}
  N
 e^{-n\n-mz}E_{ij}\otimes E_{kl}=\Big({\sum}'+{\sum}''\Big)  N
 e^{-n\n-mz}E_{ij}\otimes E_{kl}\,,
}
  \end{array}
  \eq
where the sums ${\sum}'$ and  ${\sum}''$ are defined as follows. The
total sum is over $i,j,k,l$ such that $(j,i)\in P_1$ (and $(k,l)\in
P_2$). Then the sum ${\sum}''$ is over the digonal elements $(1,N)$,
... ,$(N-1,N)$ among $(j,i)\in P_1$ (\ref{q307}), and the sum
${\sum}'$ is over the rest of the elements among $(j,i)\in P_1$ (it
is the lower triangular part of (\ref{q307})).

From (\ref{q307}) and (\ref{q308}) it follows that $j>i$ and $k>l$
for the elements in the ${\sum}'$. Moreover, $i+k=j+l$ for these
elements\footnote{Condition $i+k=j+l$ is verified directly for $n=1$
by comparing (\ref{q307}) and (\ref{q308}). To make next application
of $\tau$ one should determine $\hbox{Image}(\tau)\cap P_1\subset
P_1$, i.e. each time we return back to a subset in $P_1$. This is
why condition $i+k=j+l$ is independent of $n$.}, and $k>j$ since the
map $P_1\rightarrow P_2$ is generated by $C\times C$. Therefore,
$i<j<k$ holds true. Also, from $i=C^m_0(j)$ we have $m=j-i$. And
finally, $C^n(j)=k$, so that $n=k-j$.  In this way we showed that
the sum ${\sum}'$ provides the first term in the third line of
(\ref{q210}).

For the elements of the sum ${\sum}''$ we have $i=N>j$ and
$i+k=j+l+N$. Since $N=C_0^m(j)$ we have $j=m$. On the other hand
$C^n(N)=k$, so that $k=n$. In this way the sum ${\sum}''$ is shown
to be the first term in the last line of (\ref{q210}).

In the same way (by subdividing into two parts) the second term in
the last line of (\ref{q310}) is shown to be equal to the sum of the
second terms in the third and the fourth lines of (\ref{q210}).
$\blacksquare$


Let us comment on the origin of the general classification. It comes
from non-trivial limiting procedures (trigonometric limits)
\cite{AHZ,Burban,Smirnov} starting from the elliptic case, where the
classification is rather simple. It is based on the M. Atiyah's
classification of bundles over elliptic curves. The elliptic
$R$-matrix is fixed by its poles structure (\ref{q140}) and
quasiperiodic boundary conditions on a torus given by powers of
$N\times N$ matrices $I_1^k$, $I_2^l$ $(k,l=1,...,N-1)$, where
$I_1={\rm diag}(\exp(2\pi\imath/N),\exp(4\pi\imath/N),...,1)$ and
$(I_2)_{ij}=\e(i=j+1\,{\rm mod}\, N)$. The non-dynamical $R$-matrix
corresponds to g.c.d.$(k,N)=1$ and g.c.d.$(l,N)=1$. Otherwise,
elliptic moduli appear, which play the role of dynamical variables.



\section{Integrable tops}\label{sect3}
\setcounter{equation}{0}

Below we describe the relativistic and the non-relativistic tops
constructed by means of $R$-matrices satisfying
(\ref{q100})-(\ref{q140}). Our consideration uses results of
\cite{LOZ8,LOZ16}.
 For the relativistic models the classical
$r$-matrix structure is quadratic, while in the non-relativistic
case it is linear. In its turn the relativistic models admit two
natural (and equivalent) Lax representations: the first one includes
explicit dependence on the relativistic parameter $\eta$. It is
based on the quantum $R$-matrix. And the second one is based on the
classical $r$-matrix. The Lax pair in this description is
independent of $\eta$.

Consider a solution of the associative Yang-Baxter equation
(\ref{q100}) with the properties\footnote{In fact, it is enough
\cite{LOZ16} to have any one of  (\ref{q110}) or (\ref{q120})
conditions. In any case, we deal with $R$-matrices satisfying both
properties except the case $A_0=A_1=0$ in (\ref{q214}), where the
unitarity is degenerated.} (\ref{q110}) and (\ref{q120}) and the
following expansions near $\hbar=0$ (the classical limit):
  \beq\label{q400}
  \begin{array}{c}
      \displaystyle{
R^\h_{12}(z)=\frac{1}{\h}\,1_N\otimes 1_N+r_{12}(z)+\h\,
m_{12}(z)+O(\h^2) }
  \end{array}
  \eq
and near $z=0$
  \beq\label{q401}
  \begin{array}{c}
      \displaystyle{
R^\h_{12}(z)=\frac{1}{z}\,P_{12}+R^{\h,(0)}_{12}+zR^{\h,(1)}_{12}+O(z^2)\,,
}
  \end{array}
  \eq
  \beq\label{q402}
  \begin{array}{c}
      \displaystyle{
 R^{\h,(0)}_{12}=\frac{1}{\h}\,1_N\otimes
 1_N+r^{(0)}_{12}+O(\hbar)\,,\qquad
 r_{12}(z)=\frac{1}{z}\,P_{12}+r^{(0)}_{12}+O(z)\,.
 }
  \end{array}
  \eq
From the skew-symmetry (\ref{q110}) we have
  \beq\label{q403}
  \begin{array}{c}
  r_{12}(z)=-r_{21}(-z)\,,\qquad
  m_{12}(z)=m_{21}(-z)\,,\\ \ \\
      R^{\h,(0)}_{12}=-R^{-\h,(0)}_{21}\,,\qquad
  r_{12}^{(0)}=-r_{21}^{(0)}\,.
  \end{array}
  \eq
If the Fourier symmetry (\ref{q150}) holds true\footnote{The right
multiplication of $R$-matrix (\ref{q2}) by $P_{12}$ provides
$R_{ijkl}\rightarrow R_{ilkj}$.} as well then
  \beq\label{q404}
  \begin{array}{c}
  R^{z,(0)}_{12}=r_{12}(z)P_{12}\,,\\ \ \\
      R^{z,(1)}_{12}=m_{12}(z)P_{12}\,,\\ \ \\
  r_{12}^{(0)}=r_{12}^{(0)}P_{12}\,.
  \end{array}
  \eq
 Let us summarize the results from \cite{LOZ16}. Consider $R$-matrix, which obeys
 equations (\ref{q100})-(\ref{q140}) and has the expansions
 (\ref{q400})-(\ref{q402}). Then the Lax equations
  \beq\label{q405}
  {\dot L}(z,S)=[L(z,S),M(z,S)]
  \eq
are equivalent to equations
  \beq\label{q406}
\dot{S}=[S,J(S)]
  \eq
 in the following cases

\noindent $\bullet$ \underline{Relativistic top}:
  \beq\label{q407}
  \begin{array}{c}
     \displaystyle{
L^\eta(z,S)=\tr_2(R^\eta_{12}(z)S_2)\,,\qquad
  M^\eta(z,S)=-\tr_2(r_{12}(z)S_2)
 }
  \end{array}
  \eq
and
  \beq\label{q408}
  \begin{array}{c}
     \displaystyle{
 J^{\eta}(S)=\tr_2\Big((R^{\eta,(0)}_{12}-r_{12}^{(0)})S_2\Big)\,.
 }
  \end{array}
  \eq

\noindent $\bullet$ \underline{Non-relativistic top}:
  \beq\label{q409}
  \begin{array}{c}
     \displaystyle{
 L(z,S)=\tr_2(r_{12}(z)S_2)\,,\qquad
  M(z,S)=\tr_2(m_{12}(z)S_2)
 }
  \end{array}
  \eq
and
  \beq\label{q410}
  \begin{array}{c}
     \displaystyle{
 J(S)=\tr_2(m_{12}(0)S_2)\,.
 }
  \end{array}
  \eq
These formulae can be easily written through $R$-matrix components
(\ref{q2}). For example, the Lax matrix (\ref{q407}) is of the form
  \beq\label{q411}
  \begin{array}{c}
     \displaystyle{
 L^\eta(z,S)=\sum\limits_{i,j,k,l=1}^N R^{\eta}_{ijkl}(z)S_{lk}
 E_{ij}\,,
 }
  \end{array}
  \eq
due to $\tr(E_{kl}S)=S_{lk}$. Equivalently,
  \beq\label{q412}
  \begin{array}{c}
     \displaystyle{
 L^\eta(z,S)=\sum\limits_{i,j=1}^N L^{\eta}_{ij}(z,S)
 E_{ij}\,,\qquad
 L^{\eta}_{ij}(z,S)=\sum\limits_{k,l=1}^N
 R^{\eta}_{ijkl}(z)S_{lk}\,,
 }
  \end{array}
  \eq
and for (\ref{q408}), (\ref{q410})
  \beq\label{q4121}
  \begin{array}{l}
     \displaystyle{
 J^{\eta}(S)=\sum\limits_{i,j=1}^N E_{ij}J^\eta_{ij}(S)\,,\qquad
 J^\eta_{ij}(S)=\sum\limits_{k,l=1}^N
 (R^{\eta,(0)}_{ij,kl}-r^{(0)}_{ij,kl})S_{lk}\,,
 }
%
%
 \\ \ \\
     \displaystyle{
 J(S)=\sum\limits_{i,j=1}^N E_{ij}J_{ij}(S)\,,\qquad
 J_{ij}(S)=\sum\limits_{k,l=1}^N
 m_{ij,kl}(0)S_{lk}\,.
 }
  \end{array}
  \eq

\paragraph{Classical Sklyanin algebras and $r$-matrix structures.}
In this subsection we show that any solution of the associative
Yang-Baxter equation (\ref{q100}) with the properties provides
(\ref{q110})-(\ref{q140}) and the local expansions
(\ref{q400})-(\ref{q403}) provides the quadratic Poisson structures
of Sklyanin type. The quadratic $r$-matrix structure \cite{Skl1}
  \beq\label{q430}
  \begin{array}{c}
     \displaystyle{
c_2\{L^\eta_1(z,S),L^\eta_2(w,S)\}=[L^\eta_1(z,S)L^\eta_2(w,S),r_{12}(z-w)]\,,
 }
  \end{array}
  \eq
 where $c_2\neq 0$ is arbitrary constant, leads to the following Poisson
 brackets
  \beq\label{q431}
  \begin{array}{c}
     \displaystyle{
c_2\{S_1,S_2\}=[S_1
S_2,r_{12}^{(0)}]+[L^{\eta,(0)}_1(S)S_2,P_{12}]\,,\qquad
L^{\eta,(0)}_1(S)=\tr_3(R_{13}^{\eta,(0)}S_3)\,.
 }
  \end{array}
  \eq
for the defined above Lax matrices. These brackets are easily
obtained (see \cite{LOZ8}) by taking residues at $z=0$ and $w=0$ of
both sides of (\ref{q430}). Being written in components (\ref{q431})
takes the form:
  \beq\label{q4311}
  c_2\{S_{ij},S_{kl}\}=(L_{il}^{\n,(0)}S_{kj}-L_{kj}^{\n,(0)}S_{il})+
  \sum_{a,b=1}^{N}(S_{ia}S_{kb}r^{(0)}_{aj,bl}-r^{(0)}_{ia,kb}S_{aj}S_{bl})\,,
  \eq
 where
  \beq\label{q4312}
  L^{\eta,(0)}_{ij}=\sum\limits_{k,l=1}^N
  R^{\eta,(0)}_{ij,kl}S_{lk}\,.
  \eq
The proof of equivalence of (\ref{q431}) and (\ref{q4311}) is based
on the degeneration of (\ref{q100})
 \beq\label{q436}
  \begin{array}{c}
  \displaystyle{
 R^\hbar_{12}(x)
 R^{\hbar}_{23}(y)=R^{\hbar}_{13}(x+y)r_{12}(x)+r_{23}(y)R^\hbar_{13}(x+y)-\p_\hbar
 R_{13}^\hbar(x+y)\,,
 }
 \end{array}
 \eq
obtained by taking the limit $\eta\rightarrow\hbar$ in (\ref{q100}).

 \begin{predl}
 For the Lax
 matrix (\ref{q407}) defined by $R$-matrix satisfying the
 associative Yang-Baxter equation  (\ref{q100}) together with the
 properties (\ref{q400})-(\ref{q404}) the Poisson brackets
 (\ref{q431}) are equivalently written in the $r$-matrix form (\ref{q430}).
 \end{predl}
\noindent\underline{\em{Proof:}}\quad
 Plugging  the Lax matrix (\ref{q407})  into
 (\ref{q430}) we get the following expression for the l.h.s. of
 (\ref{q430}) up to $c_2$:\footnote{The $R$-matrices $R^\eta_{13}(z)$ and $R^\eta_{24}(w)$ commute since they are defined in different
 tensor components.}
  \beq\label{q437}
  \begin{array}{c}
     \displaystyle{
 \tr_{3,4}\Big(
 R^\eta_{13}(z)R^\eta_{24}(w)\{S_3,S_4\}\Big)\stackrel{(\ref{q431})}{=}
 \tr_{3,4}\Big(
 R^\eta_{13}(z)R^\eta_{24}(w)\Big([S_3
S_4,r_{34}^{(0)}]+[L^{\eta,(0)}_3(S)S_4,P_{34}]\Big)\Big)\,,
 }
  \end{array}
  \eq
and we are going to prove that it is equal to the r.h.s. of
(\ref{q430}):
  \beq\label{q438}
  \begin{array}{c}
     \displaystyle{
 {\rm r.h.s.}=\tr_{3,4}\Big(\Big(
 R^\eta_{13}(z)R^\eta_{24}(w)r_{12}(z-w)-r_{12}(z-w)R^\eta_{13}(z)R^\eta_{24}(w)\Big)S_3S_4\Big)\,.
 }
  \end{array}
  \eq
Let us rewrite the expression in the brackets of (\ref{q438}) using
(\ref{q436}), which we represent in the form (the skew-symmetry
(\ref{q110}) is also used)
  \beq\label{q439}
  \begin{array}{c}
     \displaystyle{
  R_{24}^\eta(w)
  r_{12}(z-w)=-R_{21}^\eta(w-z)R^\eta_{14}(z)+r_{14}(z)R_{24}^\eta(w)-\p_\eta
  R_{24}^\eta(w)
 }
  \end{array}
  \eq
 for the first term in (\ref{q438}), and
  \beq\label{q4391}
  \begin{array}{c}
     \displaystyle{
  r_{12}(z-w)R_{24}^\eta(w)
  =-R^\eta_{14}(z)R_{21}^\eta(w-z)+R_{24}^\eta(w)r_{14}(z)-\p_\eta
  R_{24}^\eta(w)
 }
  \end{array}
  \eq
 for the second one. Due to $[R^\eta_{13}(z),\p_\eta
 R^\eta_{24}(w)]=0$ we have
%
%
%
  \beq\label{q4392}
  \begin{array}{c}
     \displaystyle{
 R^\eta_{13}(z)R^\eta_{24}(w)r_{12}(z-w)-r_{12}(z-w)R^\eta_{13}(z)R^\eta_{24}(w)=
 }
 \\ \ \\
     \displaystyle{
=R^\eta_{14}(z)R_{21}^\eta(w-z)R_{13}^\eta(z)-R_{13}^\eta(z)R_{21}^\eta(w-z)R^\eta_{14}(z)+
 }
 \\ \ \\
     \displaystyle{
+R^\eta_{13}(z)r_{14}(z)R_{24}^\eta(w)-R_{24}^\eta(w)r_{14}(z)R^\eta_{13}(z)\,.
 }
  \end{array}
  \eq
 The second line of (\ref{q4392}) is cancelled out after substitution into
 (\ref{q438}) since it is skew-symmetric under renaming the numbers of the tensor components
 $3\leftrightarrow4$. Therefore, the expression (\ref{q438}) is
 simplified to
  \beq\label{q4393}
  \begin{array}{c}
     \displaystyle{
 {\rm r.h.s.}=\tr_{3,4}\Big(\Big(R^\eta_{13}(z)r_{14}(z)R_{24}^\eta(w)-R_{24}^\eta(w)r_{14}(z)R^\eta_{13}(z)\Big)S_3S_4\Big)\,.
 }
  \end{array}
  \eq
 Next, transform the latter expression using further degeneration of
 (\ref{q100}), corresponding to $z\rightarrow0$ in  (\ref{q439}) and
 (\ref{q4391}):
  \beq\label{q4394}
  \begin{array}{c}
     \displaystyle{
  R_{13}^\eta(z)r_{14}(z)=r_{34}^{(0)}R_{13}^\eta(z)+R_{14}^\eta(z)R_{43}^{\eta,(0)}
  -\p_z R_{14}^\eta(z)P_{34}+\p_\eta R_{13}^\eta(z)\,,
 }
  \end{array}
  \eq
  \beq\label{q4395}
  \begin{array}{c}
     \displaystyle{
  r_{14}(z)R_{13}^\eta(z)=R_{13}^\eta(z)r_{34}^{(0)}+R_{43}^{\eta,(0)}R_{14}^\eta(z)
  -\p_z R_{13}^\eta(z)P_{34}+\p_\eta R_{13}^\eta(z)\,.
 }
  \end{array}
  \eq
 Then the expression in the brackets of (\ref{q4393}) transforms
 into
  \beq\label{q4396}
  \begin{array}{c}
     \displaystyle{
  R^\eta_{13}(z)r_{14}(z)R_{24}^\eta(w)-R_{24}^\eta(w)r_{14}(z)R^\eta_{13}(z)=
   }
 \\ \ \\
     \displaystyle{
 =r_{34}^{(0)}R_{13}^\eta(z)R_{24}^\eta(w)-R_{24}^\eta(w)R_{13}^\eta(z)r_{34}^{(0)}+
   }
 \\ \ \\
     \displaystyle{
 +R_{14}^\eta(z)R_{43}^{\eta,(0)}R_{24}^\eta(w)-R_{24}^\eta(w)R_{43}^{\eta,(0)}R_{14}^\eta(z)+
   }
 \\ \ \\
     \displaystyle{
 +R_{24}^\eta(w)\p_z R_{13}^\eta(z)P_{34}-\p_z
 R_{14}^\eta(z)P_{34}R_{24}^\eta(w)\,.
   }
  \end{array}
  \eq
 The last line of (\ref{q4396}) vanishes being substituted into
 (\ref{q438}). Indeed, on one hand
  \beq\label{q4381}
  \begin{array}{c}
     \displaystyle{
 \tr_{3,4}\Big( \p_z R_{14}^\eta(z)P_{34}R_{24}^\eta(w)S_3S_4\Big)
 = \tr_{3,4}\Big(  P_{34}\p_z
 R_{13}^\eta(z)R_{24}^\eta(w)S_3S_4\Big)\,,
 }
  \end{array}
  \eq
 and, on the other hand,
  \beq\label{q4382}
  \begin{array}{c}
     \displaystyle{
 \tr_{3,4}\Big( R_{24}^\eta(w)\p_z R_{13}^\eta(z)P_{34} S_3S_4\Big)
 = \tr_{3,4}\Big( \p_z R_{13}^\eta(z) R_{24}^\eta(w)P_{34} S_3S_4 \Big)=
 }
 \\
      \displaystyle{
 = \tr_{3,4}\Big( \p_z R_{13}^\eta(z) R_{24}^\eta(w) S_3S_4P_{34} \Big)
 = \tr_{3,4}\Big( P_{34}\p_z R_{13}^\eta(z) R_{24}^\eta(w) S_3S_4
 \Big)\,.
 }
  \end{array}
  \eq
 The second line of (\ref{q4396}) after substitution into
 (\ref{q438}) results in the first term of the r.h.s. of  (\ref{q437}):
  \beq\label{q4383}
  \begin{array}{c}
     \displaystyle{
 \tr_{3,4}\Big( \Big( r_{34}^{(0)}R_{13}^\eta(z)R_{24}^\eta(w)-R_{24}^\eta(w)R_{13}^\eta(z)r_{34}^{(0)}
 \Big)S_3S_4\Big)=\tr_{3,4}\Big(
 R^\eta_{13}(z)R^\eta_{24}(w)\Big([S_3
S_4,r_{34}^{(0)}]\Big)\Big)\,.
 }
  \end{array}
  \eq
 Finally, the third line of (\ref{q4396}) after substitution into
 (\ref{q438}) results in the second term of the r.h.s. of (\ref{q437}):
  \beq\label{q4384}
  \begin{array}{c}
     \displaystyle{
 \tr_{3,4}\Big( \Big(
 R_{14}^\eta(z)R_{43}^{\eta,(0)}R_{24}^\eta(w)-R_{24}^\eta(w)R_{43}^{\eta,(0)}R_{14}^\eta(z)
 \Big)S_3S_4\Big)=
 }
 \\ \ \\
     \displaystyle{
 =
 \tr_{3,4}\Big(
 R^\eta_{13}(z)R^\eta_{24}(w)\Big(L^{\eta,(0)}_3(S)S_4P_{34}-P_{34}L^{\eta,(0)}_3(S)S_4\Big)\Big)\,.
 }
  \end{array}
  \eq
 The latter equality is verified as follows. Let us show that the
 first terms in the upper and lower lines of  (\ref{q4384}) are
 equal to each other (the equality of the second terms is verified
 similarly):
  \beq\label{q4385}
  \begin{array}{c}
     \displaystyle{
 \tr_{3,4}\Big( R^\eta_{13}(z)R^\eta_{24}(w)L^{\eta,(0)}_3(S)S_4P_{34} \Big)=
  \tr_{3,4}\Big( R^\eta_{13}(z)L^{\eta,(0)}_3(S)S_4P_{34}
  R^\eta_{24}(w)\Big)=
 }
 \\ \ \\
     \displaystyle{
 =\tr_{3,4,5}\Big( R^\eta_{13}(z)R_{35}^{\eta,(0)}S_5S_4P_{34}R^\eta_{24}(w) \Big)=
 \tr_{3,4,5}\Big(P_{34} R^\eta_{14}(z)R_{45}^{\eta,(0)}S_5S_3R^\eta_{24}(w) \Big)=
 }
  \\ \ \\
     \displaystyle{
 =\tr_{3,4,5}\Big(P_{34} R^\eta_{14}(z)R_{45}^{\eta,(0)}R^\eta_{24}(w) S_5S_3\Big)=
 \tr_{3,4,5}\Big( R^\eta_{14}(z)R_{45}^{\eta,(0)}R^\eta_{24}(w)
 S_5S_3P_{34}\Big)=
 }
   \\ \ \\
     \displaystyle{
 =
 \tr_{3,4,5}\Big( R^\eta_{14}(z)R_{45}^{\eta,(0)}R^\eta_{24}(w)
 S_5P_{34}S_4\Big)\,.
 }
  \end{array}
  \eq
 The last step is to take the trace over the third tensor
 component (then $P_{34}$ vanishes), and rename the component $5\leftrightarrow
 3$.

 To summarize: we
 deduce the $r$-matrix structure
 (\ref{q430}) from the brackets (\ref{q431}).
The converse statement (when the brackets (\ref{q431}) are derived
from the $r$-matrix structure(\ref{q430})) follows from the local
behavior (\ref{q401}). Indeed, by now we have proved that the
$r$-matrix structure (\ref{q430}) is equivalent to the condition
 $\tr_{3,4}\Big(
 R^\eta_{13}(z)R^\eta_{24}(w)A_{34}\Big)=0$ with $A_{12}=c_2\{S_1,S_2\}-[S_1
S_2,r_{12}^{(0)}]-[L^{\eta,(0)}_1(S)S_2,P_{12}]$, which is the
difference between l.h.s. and r.h.s. of (\ref{q431}). In order to
 prove that $A_{12}=0$ consider the expression ${\mathcal
A}_{12}(z,w)=\tr_{3,4}\Big(
 R^\eta_{13}(z)R^\eta_{24}(w)A_{34}\Big)$ locally near $z=0$ and
 $w=0$. Then form (\ref{q401}) we have ${\mathcal
A}_{12}(z,w)=z^{-1}w^{-1}\tr_{3,4}\Big(
 P_{13}P_{24}A_{34}\Big)+...=z^{-1}w^{-1}A_{12}+...$. Therefore,
 $A_{12}=0$ follows from ${\mathcal
A}_{12}(z,w)=0$.
%
%
%
%
%
%
$\blacksquare$




%
%
In the non-relativistic limit we are left with the linear $r$-matrix
structure
  \beq\label{q432}
  \begin{array}{c}
     \displaystyle{
  c_1\{L_1(z,S),L_2(w,S)\}=[L_1(z,S)+L_2(w,S),r_{12}(z-w)]\,,
 }
  \end{array}
  \eq
 which provides the Poisson-Lie brackets on ${\rm gl}_N^*$ Lie coalgebra ($c_1\neq 0$ is an arbitrary constant):
  \beq\label{q433}
  \begin{array}{c}
     \displaystyle{
 c_1\{S_1,S_2\}=[S_2,P_{12}]
 }
  \end{array}
  \eq
or
  \beq\label{q43399}
  \begin{array}{c}
     \displaystyle{
 c_1\{S_{ij},S_{kl}\}=S_{kj}\delta_{il}-S_{il}\delta_{kj}\,.
 }
  \end{array}
  \eq

The Poisson structures (\ref{q430})-(\ref{q431}) and
(\ref{q432})-(\ref{q433}) provide the Hamiltonians generating the
Euler-Arnold equations (\ref{q406}). In the relativistic case the
Hamiltonian is given by
  \beq\label{q43955}
  \begin{array}{c}
     \displaystyle{
 H^{\rm rel}=\frac{1}{c_2}\,\tr(S)\,,
 }
  \end{array}
  \eq
and for the non-relativistic case we have
  \beq\label{q43956}
  \begin{array}{c}
     \displaystyle{
 H^{\rm non-rel}=\frac{1}{2c_1}\,\tr(SJ(S))\,.
 }
  \end{array}
  \eq
 In the relativistic case the Hamiltonian is linear, while the
 Poisson structure is quadratic (in variables $S$), and  vice versa
 for non-relativistic models.

\subsection{The case of non-standard $R$-matrix}
In order to describe the tops explicitly it is enough to write down
all $R$-matrices and related coefficients of expansions entering
(\ref{q407})-(\ref{q4121}). Below is the summary based on the
$R$-matrix (\ref{q2101}):
  \beq\label{q21011}
   \begin{array}{c}
   \displaystyle{
  R^{\n}_{ij,kl}(z)= \de_{ij}\de_{kl}\de_{ik}\,\frac{N}{2}\,\Big(\coth(Nz/2)+\coth(N\n/2)\Big)+
  }
  \\ \ \\
   \displaystyle{
  +\, \de_{ij}\de_{kl}\e(i\ne k)\frac{N e^{(i-k)\n-\sgn(i-k)N\n/2}}{2\sinh(N\n/2)}+\de_{il}\de_{kj}
  \e(i\ne k)\frac{N e^{(i-k)z-\sgn(i-k)Nz/2}}{2\sinh(Nz/2)}\, +
  }
  \\ \ \\
     \displaystyle{
  + N\de_{i+k,j+l}e^{(i-j)z+(j-k)\n}\Big(\e(i{<}j{<}k) -\e(k{<}j{<}i)
  \Big)+
  }
  \\ \ \\
     \displaystyle{
  + N
  e^{-N\Lambda}
  \de_{i+k,j+l+N}\Big(\de_{iN}e^{-jz-l\n}-\de_{kN}e^{lz+j\n}\Big)\,.
 }
  \end{array}
  \eq
The classical $r$-matrix:
  \beq\label{q440}
   \begin{array}{c}
   \displaystyle{
  r_{ij,kl}(z)= \de_{ij}\de_{kl}\de_{ik}\,\frac{N}{2}\,\coth(Nz/2)+
  }
  \\ \ \\
   \displaystyle{
  +\, \de_{ij}\de_{kl}\e(i\ne k) \Big((i-k)-\frac{N\sgn(i-k)}2\Big)
  +\de_{il}\de_{kj}
  \e(i\ne k)\frac{N e^{(i-k)z-\sgn(i-k)Nz/2}}{2\sinh(Nz/2)}\, +
  }
  \\ \ \\
     \displaystyle{
  + Ne^{(i-j)z}\de_{i+k,j+l}\Big(\e(i{<}j{<}k) -\e(k{<}j{<}i)
  \Big)
  + N
  e^{-N\Lambda}
  \de_{i+k,j+l+N}\Big(e^{-jz}\de_{iN}-e^{lz}\de_{kN}\Big)\,.
 }
  \end{array}
  \eq
The next coefficient in the expansion (\ref{q400}):
  \beq\label{q441}
   \begin{array}{c}
   \displaystyle{
  m_{ij,kl}(z)= \de_{ij}\de_{kl}\de_{ik}\,\frac{N^2}{12}+
  \, \de_{ij}\de_{kl}\e(i\ne k) \Big(\frac{(i-k)^2}{2}-\frac{N^2}{12}-\frac{N}{2}|i-k|\Big)
   +
  }\\ \
  \end{array}
  \eq
$$
   \displaystyle{
  + N(j-k)e^{(i-j)z}\de_{i+k,j+l}\Big(\e(i{<}j{<}k) -\e(k{<}j{<}i)
  \Big)
  - N
  e^{-N\Lambda}
  \de_{i+k,j+l+N}\Big(l\,e^{-jz}\de_{iN}+j\,e^{lz}\de_{kN}\Big)\,.
   }
$$
Its value at $z=0$ entering the inverse inertia tensor in the
non-relativistic case (\ref{q410}) or (\ref{q4121}):
  \beq\label{q4411}
   \begin{array}{c}
   \displaystyle{
  m_{ij,kl}(0)= \de_{ij}\de_{kl}\de_{ik}\,\frac{N^2}{12}+
  \, \de_{ij}\de_{kl}\e(i\ne k) \Big(\frac{(i-k)^2}{2}-\frac{N^2}{12}-\frac{N}{2}|i-k|\Big)
   +
  }\\ \
  \end{array}
  \eq
$$
   \displaystyle{
  + N(j-k)\de_{i+k,j+l}\Big(\e(i{<}j{<}k) -\e(k{<}j{<}i)
  \Big)
  - N
  e^{-N\Lambda}
  \de_{i+k,j+l+N}\Big(l\,\de_{iN}+j\,\de_{kN}\Big)\,.
   }
$$
The coefficient from the expansions (\ref{q401}) and (\ref{q402})
entering the relativistic inverse inertia tensor (\ref{q408}) or
(\ref{q4121}):
  \beq\label{q442}
   \begin{array}{c}
   \displaystyle{
  R^{\eta,(0)}_{ij,kl}=r_{ilkj}(\eta)= \de_{ij}\de_{kl}\de_{ik}\,\frac{N}{2}\coth(N\n/2)+
  }
  \\ \ \\
   \displaystyle{
  +\, \de_{ij}\de_{kl}\e(i\ne k)\frac{N e^{(i-k)\n-\sgn(i-k)N\n/2}}{2\sinh(N\n/2)}+\de_{il}\de_{kj}
  \e(i\ne k) \Big((i-k)-\frac{N\sgn(i-k)}2\Big) +
  }
  \\ \ \\
     \displaystyle{
  + Ne^{(j-k)\n}\,\de_{i+k,j+l}\Big(\e(i{<}j{<}k) -\e(k{<}j{<}i)
  \Big)
  + N
  e^{-N\Lambda}
  \de_{i+k,j+l+N}\Big(e^{-l\n}\de_{iN}-e^{j\n}\de_{kN}\Big)
 }
  \end{array}
  \eq
and
  \beq\label{q443}
   \begin{array}{c}
   \displaystyle{
  r_{ij,kl}^{(0)}=
\Big(\de_{ij}\de_{kl}\e(i\ne k)  +\de_{il}\de_{kj}
  \e(i\ne k)\Big)\Big((i-k)-\frac{N\sgn(i-k)}2\Big)+
  }
  \\ \ \\
     \displaystyle{
  + N\de_{i+k,j+l}\Big(\e(i{<}j{<}k) -\e(k{<}j{<}i)
  \Big)
  + N
  e^{-N\Lambda}
  \de_{i+k,j+l+N}\Big(\de_{iN}-\de_{kN}\Big)\,.
 }
  \end{array}
  \eq

\paragraph{Lax pairs.} The Lax matrix of the relativistic top constructed by means of (\ref{q21011}) is of the following form. For
$i=j$:
  \beq\label{q4431}
   \begin{array}{c}
   \displaystyle{
L_{ii}^{\n}(z)=\frac{N}{2}\big(\coth(Nz/2)+\coth(N\n/2)\big)S_{ii}+
 }
 \\
   \displaystyle{
  +\frac{N}{2\sinh(N\n/2)}\left(e^{-N\n/2}\sum_{k=1}^{i-1}e^{(i-k)\n}S_{kk}+e^{N\n/2}\sum_{k=i+1}^{N}e^{(i-k)\n}S_{kk}\right)\,,
 }
  \end{array}
  \eq
for $i<j$:
  \beq\label{q4432}
   \begin{array}{c}
   \displaystyle{
L_{ij}^{\n}(z)=\frac{N\exp(Nz/2+(i-j)z)}{2\sinh(Nz/2)}S_{ij}+N\sum_{k=j+1}^{N}e^{(i-j)z+(j-k)\n}S_{i-j+k,k}\,,
 }
  \end{array}
  \eq
and for $i>j$:
  \beq\label{q4433}
   \begin{array}{c}
   \displaystyle{
L_{ij}^{\n}(z)=\frac{N\exp(-Nz/2+(i-j)z)}{2\sinh(Nz/2)}S_{ij}-N\sum_{k=1}^{j-1}e^{(i-j)z+(j-k)\n}S_{i-j+k,k}-
 }
 \\
   \displaystyle{
  -Ne^{-N\Lambda}
  e^{(i-j)z+j\n}S_{i-j,N}+\de_{iN}Ne^{-N\Lambda}\sum_{k=j+1}^{N}e^{-jz+(j-k)\n}S_{k-j,k}\,.
 }
  \end{array}
  \eq
 From the definitions (\ref{q407}), (\ref{q409}) and the
expansion (\ref{q400}) it follows that
  \beq\label{q4435}
   \begin{array}{c}
   \displaystyle{
-M^\eta(z)=L(z)=\res\limits_{\eta=0}\Big(\eta^{-1}L^{\eta}(z)\Big)\,,
\qquad M(z)=\res\limits_{\eta=0}\Big(\eta^{-2}L^{\eta}(z)\Big)\,.
 }
  \end{array}
  \eq
Similarly, the expansion (\ref{q401}) near $z=0$ yields
  \beq\label{q4437}
   \begin{array}{c}
   \displaystyle{
L^\eta(z)=\frac{1}{z}S+L^{\eta,(0)}(S)+O(z)\,,\qquad
L^{\eta,(0)}(S)=\tr_2\Big(R_{12}^{\eta,(0)}S_2\Big)=\res\limits_{z=0}\Big(z^{-1}L^{\eta}(z)\Big)\,.
 }
  \end{array}
  \eq

\paragraph{Example: $GL_2$ top.} In this case we deal with the
following quantum
  \beq\label{q444}
   \begin{array}{c}
  R^\hbar(z)=\left(\begin{array}{cccc} \coth(z)+\coth(\hbar) & 0 & 0 & 0\vphantom{\Big|}
  \\ 0 & \sinh^{-1}(\hbar) & \sinh^{-1}(z) & 0\vphantom{\Big|}
  \\ 0 & \sinh^{-1}(z) & \sinh^{-1}(\hbar) & 0\vphantom{\Big|}
  \\ -4\,e^{-2\Lambda}\sinh(z+\hbar) & 0 & 0 & \coth(z)+\coth(\hbar)          \vphantom{\Big|} \end{array} \right)
  \end{array}
  \eq
and classical
  \beq\label{q445}
   \begin{array}{c}
  r(z)=\left(\begin{array}{cccc} \coth(z)&0&0&0\\0&0&\sinh^{-1}(z)&0
  \\0&\sinh^{-1}(z)&0&0\\-4\,e^{-2\Lambda}\sinh(z)&0&0&\coth(z) \end{array} \right)
  \end{array}
  \eq
$R$-matrices.
In the relativistic case this provides the Lax pair
  \beq\label{q446}
   \begin{array}{c}
L^\eta(z,S) = \left(\begin{array}{cc}
   \displaystyle{ S_{11}\Big(\coth(z)+\coth(\eta)\Big)+\frac{S_{22}}{\sinh(\eta)} } &
   \displaystyle{ \frac{S_{12}}{\sinh(z)} }
\\ \
\\    \displaystyle{ \frac{S_{21}}{\sinh(z)}-4e^{-2\Lambda}S_{12}\sinh(z+\eta) }&
   \displaystyle{ S_{22}\Big(\coth(z)+\coth(\eta)\Big)+\frac{S_{11}}{\sinh(\eta)} }
\end{array} \right)
  \end{array}
  \eq
  \beq\label{q447}
   \begin{array}{c}
M^\n(z,S)= -\left(\begin{array}{cc}
   \displaystyle{ \coth(z)S_{11} } &
   \displaystyle{ \frac{\textstyle S_{12}}{\textstyle\sinh(z)}}
\\ \
\\    \displaystyle{ \frac{\textstyle
S_{21}}{\textstyle\sinh(z)}-4e^{-2\Lambda}\sinh(z)S_{12} }&
   \displaystyle{ \coth(z)S_{22} }
\end{array} \right)
  \end{array}
  \eq
and the inverse inertia tensor
  \beq\label{q448}
   \begin{array}{c}
J^\n(S)=\left(\begin{array}{cc} \coth(\eta)S_{11}+\frac{\textstyle
S_{22}}{\textstyle\sinh(\eta)} & 0 \\ \ \\
-4e^{-2\Lambda}\sinh(\eta)S_{12} & \frac{\textstyle
S_{11}}{\textstyle\sinh(\eta)}+\coth(\eta)S_{22}
\end{array}\right)
  \end{array}
  \eq
In the non-relativistic case the Lax matrix is defined by
(\ref{q447}): $L(z,S)=-M^\n(z,S)$. The accompany matrix is as
follows
  \beq\label{q449}
   \begin{array}{c}
      \displaystyle{
M(z,S)=\frac{1}{6}\left(\begin{array}{cc} 2S_{11}-S_{22} & 0 \\ \ \\
-24\,e^{-2\Lambda}\cosh(z)S_{12} & -S_{11}+2S_{22}\end{array}\right)
  }
  \end{array}
  \eq
The inverse inertia tensor acquires the form:
  \beq\label{q450}
   \begin{array}{c}
      \displaystyle{
J(S)=\frac{1}{6}\left(\begin{array}{cc} 2S_{11}-S_{22} & 0 \\ \ \\
-24\,e^{-2\Lambda}S_{12} & -S_{11}+2S_{22}\end{array}\right)
  }
  \end{array}
  \eq


\noindent $\bullet$ \underline{Relativistic top ($\eta$-independent
description)}:

Another one description for the relativistic top is available, which
is similar to original construction \cite{Skl1}.
Instead of usage of the quantum $R$-matrix (\ref{q407}) consider the
traceless part of the non-relativistic Lax matrix and supplement it
by  the scalar term $s_01_N$:
  \beq\label{q413}
  \begin{array}{c}
     \displaystyle{
\tL(z,S)=s_01_N+L(z,S)-\frac{1_N}{N}\,\tr L(z,S)\,,\qquad
s_0=\frac{\tr S}{N}\,,
 }
  \end{array}
  \eq
where $s_0$ is a dynamical variable. In fact, it is the Hamiltonian
since $\tr\tL=Ns_0$. The Lax equations do not change because
$L(z,S)$ and $\tL(z,S)$ differ from each other by only a scalar
matrix. So that the $M$-matrix for (\ref{q413}) is the same as in
(\ref{q407}). However, the Poisson structures are different (see
below). It happens because of the bihamiltonian structure in this
kind of models \cite{KLO,LOZ8}.

As was mention in \cite{LOZ8} (see also \cite{CLOZ}) there is a
relation between  the Lax matrices (\ref{q407}) and (\ref{q413}).
Similarly, to the rational case we have
  \beq\label{q414}
  \begin{array}{c}
     \displaystyle{
 L^\n(z-\frac{\n}{N},\tL(\frac{\n}{N},S))=\frac{\tr(L^\n(z-\frac\eta{N},S))}{\tr(S)}\tL(z,S)\,.
 }
  \end{array}
  \eq
These relation can be verified directly using explicit formulae
(\ref{q4431})-(\ref{q4433}).

The quadratic Poisson structure takes the form
  \beq\label{q434}
  \begin{array}{c}
     \displaystyle{
  \{\tL_1(z,S),\tL_2(w,S)\}=\frac{1}{c_2}\,[\tL_1(z,S)\tL_2(w,S),r_{12}(z-w)]\,,
 }
  \end{array}
  \eq
 and provides the following Poisson brackets:
  \beq\label{q435}
  \begin{array}{c}
     \displaystyle{
c_2
\{S_1,S_2\}=s_0[S_2,P_{12}]+[S_1S_2,r^{(0)}_{12}]+[\tr_3(r_{13}^{(0)}S_3)S_2,P_{12}]\,.
 }
  \end{array}
  \eq
The latter is verified similarly to the $\eta$-dependent case
(\ref{q430})-(\ref{q431}).

\subsection{The case of general $R$-matrix}

The summary of the integrable tops data in the general case is based
on the expansions of the $R$-matrix (\ref{q302}):
  \beq\label{q500}
   \begin{array}{c}
      \displaystyle{
  R^{\eta}_{12}(z)=\frac{N}{2}\Big(\coth(Nz/2)+\coth(N\n/2)\Big)\sum_{i} E_{ii}\otimes
  E_{ii}+
  }
   \\ \ \\
   \displaystyle{
  +\frac{N}{e^{N\n}-1}\sum\limits_{0<n<N\,,\, i=C^n(k)} e^{n\n}E_{ii}\otimes E_{kk}+
  \frac{N}{e^{Nz}-1}\sum\limits_{0<m<N\,,\,k=C^m_0(i)}   e^{mz} E_{ik}\otimes
  E_{ki}+
  }
  \\ \ \\
   \displaystyle{
  +\sum\limits_{\hbox{\tiny${\begin{array}{c}0<m<N\,,\,n>0,\\
  i=C^m_0(j)\,,\,\tau^n(j,i)=(k,l)\end{array}}$}}
  N
  \left(e^{-n\n-mz}E_{ij}\otimes E_{kl}- e^{n\n+mz}E_{kl}\otimes
  E_{ij}\right)\,,
   }
  \end{array}
  \eq
The classical $r$-matrix and the next coefficient of the classical
limit (\ref{q400}) are as follows:
  \beq\label{q501}
   \begin{array}{c}
      \displaystyle{
  r_{12}(z)=\frac{N}{2}\coth(Nz/2)\sum_{i} E_{ii}\otimes
  E_{ii}+
  }
   \\ \ \\
   \displaystyle{
  +\sum\limits_{0<n<N\,,\, i=C^n(k)} \Big(n-\frac{N}{2}\Big)E_{ii}\otimes E_{kk}+
  \frac{N}{e^{Nz}-1}\sum\limits_{0<m<N\,,\,k=C^m_0(i)}   e^{mz} E_{ik}\otimes
  E_{ki}+
  }
  \\ \ \\
   \displaystyle{
  +\sum\limits_{\hbox{\tiny${\begin{array}{c}0<m<N\,,\,n>0,\\
  i=C^m_0(j)\,,\,\tau^n(j,i)=(k,l)\end{array}}$}}
  N
  \left(e^{-mz}E_{ij}\otimes E_{kl}- e^{mz}E_{kl}\otimes
  E_{ij}\right)
   }
  \end{array}
  \eq
and
  \beq\label{q502}
   \begin{array}{c}
      \displaystyle{
  m_{12}(z)=\frac{N^2}{12}\sum_{i} E_{ii}\otimes
  E_{ii}
  +\frac{1}{12}\sum\limits_{0<n<N\,,\, i=C^n(k)} \Big(6n^2-6nN+N^2\Big)E_{ii}\otimes
  E_{kk}-
  }
  \\ \ \\
   \displaystyle{
  -\sum\limits_{\hbox{\tiny${\begin{array}{c}0<m<N\,,\,n>0,\\
  i=C^m_0(j)\,,\,\tau^n(j,i)=(k,l)\end{array}}$}}
  Nn
  \left(e^{-mz}E_{ij}\otimes E_{kl}+ e^{mz}E_{kl}\otimes
  E_{ij}\right)\,.
   }
  \end{array}
  \eq
The first non-trivial coefficients from the expansions (\ref{q401}),
(\ref{q402}) are of the form:
  \beq\label{q503}
   \begin{array}{c}
      \displaystyle{
  R^{\eta,(0)}_{12}=\frac{N}{2}\coth(N\n/2)\sum_{i} E_{ii}\otimes
  E_{ii}+
  }
   \\ \ \\
   \displaystyle{
  +\frac{N}{e^{N\n}-1}\sum\limits_{0<n<N\,,\, i=C^n(k)} e^{n\n}E_{ii}\otimes E_{kk}+
  \sum\limits_{0<m<N\,,\,k=C^m_0(i)}   \Big(m-\frac{N}{2}\Big) E_{ik}\otimes
  E_{ki}+
  }
  \\ \ \\
   \displaystyle{
  +\sum\limits_{\hbox{\tiny${\begin{array}{c}0<m<N\,,\,n>0,\\
  i=C^m_0(j)\,,\,\tau^n(j,i)=(k,l)\end{array}}$}}
  N
  \left(e^{-n\n}E_{ij}\otimes E_{kl}- e^{n\n}E_{kl}\otimes
  E_{ij}\right)
   }
  \end{array}
  \eq
and
  \beq\label{q504}
   \begin{array}{c}
      \displaystyle{
  r^{(0)}_{12}=
  \sum\limits_{0<n<N\,,\, i=C^n(k)} \Big(n-\frac{N}{2}\Big)E_{ii}\otimes E_{kk}+
  \sum\limits_{0<m<N\,,\,k=C^m_0(i)}   \Big(m-\frac{N}{2}\Big) E_{ik}\otimes
  E_{ki}+
  }
  \\ \ \\
   \displaystyle{
  +\sum\limits_{\hbox{\tiny${\begin{array}{c}0<m<N\,,\,n>0,\\
  i=C^m_0(j)\,,\,\tau^n(j,i)=(k,l)\end{array}}$}}
  N
  \left(E_{ij}\otimes E_{kl}- E_{kl}\otimes
  E_{ij}\right)\,.
   }
  \end{array}
  \eq


\section{Relation to Ruijsenaars-Schneider model}\label{sect4}
\setcounter{equation}{0}



Introduce the matrix \cite{AHZ}\footnote{It is the intertwining
matrix relating the non-standard $R$-matrix and the trigonometric
Felder's dynamical $R$-matrix through the quantum IRF-Vertex
correspondence.}
%
%
  \beq\label{q700}
   \begin{array}{c}
      \displaystyle{
 g(z, q)=\Xi(z,q)D^{-1}(q)\in\Mat\,,
   }
  \end{array}
  \eq
 where
  \beq\label{q701}
   \begin{array}{c}
      \displaystyle{
 \Xi_{ij}(z,q)=e^{(i-1)(z-{\bar q}_j)}+(-1)^{N}e^{-(z-{\bar q}_j)}\de_{iN}
   }
  \end{array}
  \eq
and
  \beq\label{q702}
   \begin{array}{c}
      \displaystyle{
   D_{ij}( q)=\de_{ij}
  \prod\limits_{k\neq
 i}\Big(e^{-{\bar q}_i}-e^{-{\bar q}_k}\Big)\,.
   }
  \end{array}
  \eq
The matrices depend on $z$ and the set of variables $q_1,...,q_N$.
The variables ${\bar q}_1,...,{\bar q}_N$ are obtained by transition
to the center of mass frame:
  \beq\label{q7021}
   \begin{array}{c}
      \displaystyle{
 {\bar q}_i=q_i-\frac1N\sum\limits_{k=1}^Nq_k\,.
   }
  \end{array}
  \eq
The determinant of the matrix $\Xi$ is as follows:
  \beq\label{q7022}
   \begin{array}{c}
      \displaystyle{
 \det\Xi(z,q)=e^{zN(N-1)/2}(1-e^{-Nz})\prod\limits_{i>j}^N\left(e^{-{\bar q}_i}-e^{-{\bar
 q}_j}\right)\,.
   }
  \end{array}
  \eq
That is $\Xi(z,q)$ is degenerated at $z=0$.

Our statement is that the following matrix
  \beq\label{q7023}
   \begin{array}{c}
      \displaystyle{
{ L}^{\rm RS}(z)= g^{-1}(z) g(z+\n)e^{P/c}\,,\qquad P={\rm
diag}(p_1,p_2,...,p_N)
   }
  \end{array}
  \eq
is the Lax matrix of the trigonometric Ruijsenaars-Schneider model.
More precisely,
  \beq\label{q706}
   \begin{array}{c}
      \displaystyle{
L^{\rm RS}_{ij}(z)
  =e^{\frac{N-2}{2}\,\eta}\sinh(\eta/2)
  \left(\coth\left(\frac{Nz}{2}\right)+\coth\left(\frac{q_i-q_j+\n}{2}\right)\right)e^{p_j/c}
  \prod^N_{k\ne
  j}\frac{\sinh\left(\frac{q_j-q_k-\n}{2}\right)}{\sinh\left(\frac{q_j-q_k}{2}\right)}\,.
   }
  \end{array}
  \eq
The proof is obtained by direct verification, which is similar to
calculations performed in \cite{AASZ,LOZ8} in the rational case. One
should introduce the set of elementary symmetric polynomials
$\sigma_k(q)$
 of $N$ variables $\{e^{-{\bar q}_1},...,e^{-{\bar q}_N}\}$
  \beq\label{q703}
   \begin{array}{c}
      \displaystyle{
  \prod_{k=1}^N(\zeta-e^{-{\bar q}_k})=\sum_{k=0}^N(-1)^k\zeta^k\sigma_k(q)
   }
  \end{array}
  \eq
 and $N$ sets of similar functions $\{\check{\sigma}_{k,i}(q),i=1,...,N\}$ defined for the
 sets $\{e^{-{\bar q}_1},...,e^{-{\bar q}_N}\}\backslash \{e^{-{\bar
 q}_i}\}$ of $N-1$ variables  each:
  \beq\label{q7031}
   \begin{array}{c}
      \displaystyle{
  \prod_{k\neq i}^N(\zeta-e^{-{\bar
  q}_k})=\sum_{k=0}^N(-1)^k\zeta^k\check{\sigma}_{k,i}(q)\,.
   }
  \end{array}
  \eq
The inverse of $\Xi$ is then written as follows:
  \beq\label{q705}
   \begin{array}{c}
      \displaystyle{
 (\Xi^{-1})_{ij}(z, q)=\frac{(-1)^{j-1}e^{(N-j+1)z}}{e^{Nz}-1}
 \frac{\Big(\check{\sigma}_{j-1,i}(q)+e^{-{\bar q}_i}\check{\sigma}_{j,i}(q)e^{-Nz}\Big)}{\prod\limits_{k\neq
 i}\Big(e^{-{\bar q}_i}-e^{-{\bar q}_k}\Big)}\,.
   }
  \end{array}
  \eq

Consider the gauge transformed Lax matrix
  \beq\label{q707}
   \begin{array}{c}
      \displaystyle{
L^\n(z)=g(z){\ti L}^{RS}(z)g^{-1}(z)=g(z+\n)e^{P/c}g^{-1}(z)
   }
  \end{array}
  \eq
Then\footnote{The origins of factorization of the Lax pairs
(\ref{q7023}) and (\ref{q707}), (\ref{q708}) are discussed in
\cite{VZ}.}
  \beq\label{q708}
   \begin{array}{c}
      \displaystyle{
L^\n(z)=\tr \Big(R_{12}^\eta(z)S_2(p,q)\Big)
   }
  \end{array}
  \eq
 with the non-standard $R$-matrix (\ref{q2101}), where
 $\Lambda=\sqrt{-1}\pi$. Put it differently, the matrix (\ref{q708}) coincides with
 (\ref{q4431})-(\ref{q4433}) when $\Lambda=\sqrt{-1}\pi$.
  The change of variables is as follows:
  \beq\label{q709}
   \begin{array}{c}
      \displaystyle{
 S_{ij}(p,q)=\frac{(-1)^{j}\sigma_{j}(q)e^{(i-1)\n}}{N}\sum_{n=1}^{N}\frac{e^{p_n/c}}{\prod\limits_{k:k\ne
 n}(e^{-{\bar q}_n}-e^{-{\bar q}_k})}\left(e^{-(i-1){\bar q}_n}+\frac{(-1)^N
 \de_{iN}}{e^{N\n-{\bar q}_n}}\right)\,.
   }
  \end{array}
  \eq
The Poisson structure for $(p,q)$ variables is canonical, i.e.
  \beq\label{q710}
   \begin{array}{c}
      \displaystyle{
   \{p_i,q_j\}=\delta_{ij}\qquad \hbox{or}\qquad
   \{p_i,{\bar q}_j\}=\delta_{ij}-\frac1N\,.
   }
  \end{array}
  \eq
After some tedious calculations it can be verified that the Poisson
brackets $\{S_{ij}(p,q),S_{kl}(p,q)\}$ evaluated through
(\ref{q710}) coincide with (\ref{q4311}) with $c_2=Nc$ and
$r_{12}^{(0)}$ from (\ref{q443}). In particular, it is useful to
notice for the proof that the matrix (\ref{q709}) is of rank 1, i.e.
  \beq\label{q711}
   \begin{array}{c}
      \displaystyle{
S_{ij}(p,q)=a_i(p,q)b_j(q)\,,
   }
   \\ \ \\
      \displaystyle{
a_i(p,q)=\frac{e^{(i-1)\n}}{N}\sum_{n=1}^{N}\frac{e^{p_n/c}}{\prod\limits_{k:k\ne
 n}(e^{-{\bar q}_n}-e^{-{\bar q}_k})}\left(e^{-(i-1){\bar q}_n}+\frac{(-1)^N
 \de_{iN}}{e^{N\n-{\bar q}_n}}\right)\,,\qquad
 b_j(q)=(-1)^{j}\sigma_{j}(q)\,.
      }
  \end{array}
  \eq
 In this case $S_{ij}S_{kl}=S_{il}S_{kj}$, and the Poisson
structure (\ref{q4311}) takes the (relatively simple) form:
  \beq\label{q4436}
   \begin{array}{c}
       \displaystyle{
\{S_{ij},S_{kl}\}=\frac{1}{Nc}(L_{il}^{\n,(0)}S_{kj}-L_{kj}^{\n,(0)}S_{il})+\frac{2}{Nc}(k-i)S_{ij}S_{kl}+
 }
 \\ \ \\
    \displaystyle{
  +\frac{\e(i{>}k)}{c}\sum_{p=0}^{i-k-1}S_{i-p,j}S_{k+p,l}-\frac{\e(i{<}k)}{c}\sum_{p=0}^{k-i-1}S_{i+p,j}S_{k-p,l}+
   }
  \\ \ \\
     \displaystyle{
  +\frac{(-1)^N\de_{kN}}{c}\sum_{p=1}^{i-1}S_{i-p,j}S_{p,l}-\frac{(-1)^N\de_{iN}}{c}\sum_{p=1}^{k-1}S_{pj}S_{k-p,l}\,.
  }
    \end{array}
  \eq

\paragraph{Non-relativistic limit.} The Calogero-Moser-Sutherland models appears from the above results by taking the
non-relativistic limit, when $\eta=\nu/c$ and $c\rightarrow\infty$.
The Lax matrix arising from (\ref{q706}) is of the form\footnote{It
is easy to verify that  $p_i\rightarrow{\dot q}_i(p,q)$ with ${\dot
q}_i(p,q)$ from (\ref{q4271}) is a canonical map, i.e. $\{ {\dot
q}_i(p,q),q_j \}=\delta_{ij}$.}\footnote{Let us mention that there
is another one application of the associative Yang-Baxter equation
to the models of the Calogero-Moser-Sutherland type and related
long-range spin chains \cite{LOZ9,SeZ2}.}
    \begin{equation}
   \label{q4271}
      \begin{array}{c}
     \displaystyle{
   L^{\rm CM}_{ij}(z)=\delta_{ij}( {\dot q}_i+ \nu \coth(Nz))+   \nu(1-\delta_{ij})
\Big(\coth\Big(\frac{q_{i}-q_{j}}{2}\Big) + \coth(Nz)\Big)\,,
   }
   \\ \ \\
        \displaystyle{
 {\dot q}_i=p_{i} + \nu (N-2) -\nu \sum\limits_{k \neq i}^{N}
   \coth\Big(\frac{q_{i}-q_{j}}{2}\Big)\,.
    }
   \end{array}
   \end{equation}
Similarly, the non-relativistic top (\ref{q409}) comes from
(\ref{q408}). The gauge transformation (\ref{q707}) holds on at the
level of non-relativistic models as well \cite{LOZ,VZ}. That is
  \beq\label{q4272}
   \begin{array}{c}
      \displaystyle{
L(z)=\tr \Big(r_{12}(z)S_2(p,q)\Big)=g(z)L^{\rm CM}(z)g^{-1}(z)\,.
   }
  \end{array}
  \eq
The residue of both parts of the latter relation provides explicit
change of variables, or the non-relativistic limit of (\ref{q711}):
  \beq\label{q4273}
   \begin{array}{c}
      \displaystyle{
S_{ij}(p,q)=a_i(p,q)b_j(q)\,,\qquad b_j(q)=(-1)^{j}\sigma_{j}(q)\,,
   }
   \\ \ \\
      \displaystyle{
a_i(p,q)=\frac{1}{N}\sum_{n=1}^{N}\frac{\left(p_n+(i-1)\nu\right)\left(e^{-(i-1){\bar
q}_n}+(-1)^N
 \de_{iN}e^{{\bar q}_n}\right)-N\nu(-1)^N
 \de_{iN}e^{{\bar q}_n}}{\prod\limits_{k:k\ne
 n}(e^{-{\bar q}_n}-e^{-{\bar q}_k})}\,.
      }
  \end{array}
  \eq
The Poisson brackets $\{S_{ij}(p,q),S_{kl}(p,q)\}$ computed via the
canonical structure (\ref{q710}) reproduce (\ref{q43399}) with
$c_1=N$, and the value of the Casimir functions are given by the
powers of the
Calogero-Moser-Sutherland coupling constant 
  \beq\label{q4274}
   \begin{array}{c}
      \displaystyle{
\tr(S^k)=\nu^k\,.
   }
  \end{array}
  \eq
 Thus, the Calogero-Moser-Sutherland model is gauge equivalent to
 the non-relativistic top with special
 values of the Casimir functions corresponding to the coadjoint orbit (of ${\rm GL}_N$ group) of minimal
 dimension. Apart from the gauge transformation we obtain explicit
 change of variables (in fact, a canonical map) $(p_i,q_j)\rightarrow(a_i(p,q),b_i(q))$, where
 $b_i$ are elementary symmetric functions. These variables are known
 in
 the quantum Calogero-Moser-Sutherland model \cite{Perelomov}.


\begin{small}
 
\end{small}

\end{document}